# Recommendations for Planning Inclusive Astronomy Conferences

Originally written by the Inclusive Astronomy 2 Local Organizing Committee: Brian Brooks, Keira Brooks, Lea Hagen, Nimish Hathi, Samantha Hoffman, James Paranilam, and Laura Prichard

July 2020

## Contents

# Introduction

## About the Inclusive Astronomy Conference Series

The Inclusive Astronomy (IA) conference series aims to create a safe space where community members can listen to the experiences of marginalized individuals in astronomy, discuss actions being taken to address inequities, and give recommendations to the community for how to improve diversity, equity, and inclusion in astronomy. The first IA (hereafter IA1) was held in Nashville, TN, USA, 17–19 June, 2015. The [Inclusive Astronomy 2 (IA2)](#) conference was held in Baltimore, MD, USA, 14–15 October, 2019.

## This Document

*Note: Throughout this document, "we" will refer to the Local Organizing Committee (LOC) for IA2.*

When planning an IA conference, there are many things to keep in mind and plan for. What we lay out below is a detailed document with our recommendations for leading and organizing such a conference, based on IA1 recommendations, our experiences, feedback from the community, and feedback from the attendees of IA2. We do not assume that we have all the answers nor that we did everything "correctly," but instead give recommendations based on our gained experience. This document includes details that may not be applicable to all conferences, however, we anticipate it will be relevant to many organizing committees striving to make their conferences more inclusive, regardless of whether inclusivity is their primary focus. Our hope for this document is for it to evolve and be built upon for future organizers of IA and related conferences.

At the highest level, we have found that this conference takes a significant time and emotional commitment. It is likely that the organizers will be taking on tasks that are rarely done at large astronomy conferences, which means that some activities may take longer to organize and execute than expected. We recommend that ample time is given to plan the conference, clear goals be laid out at the very beginning of planning, different needs (such as access to information and accessibility of the conference) be addressed at every stage of planning, and there be point people on the organizing committee that are focused on specific conference goals.

From the feedback we received — in the planning stage, during, and after the conference — it was clear that all participants have different needs and priorities, some of which are incompatible with those of other participants. We have, as much as possible, included that feedback in these recommendations. At times there was no majority, and we attempt to present different options for those situations.

## Top Recommendations

**These are our top-level recommendations that we view as the most important.**
- Start 1–1.5 years in advance. We cannot stress enough how important it is to give yourself significant time to plan this conference.
- Seek involvement from people of many identities and experts in different areas in the planning of the conference.
- Be transparent in decision-making, and keep records of how decisions are made (e.g. meeting minutes, progress checklists).
- Utilize resources from outside the astronomy community.
- Accessibility and access needs must be addressed at every stage in the process.
- The organizing committees should carefully set up planning structures to ensure success. This includes finding ways to enable good communication and instituting multi-level leadership.

You will not make everyone happy and will likely not be able to address every need. However, you can ensure that you are meeting the primary needs of different groups by requiring a diverse organizing committee and consulting experts that are sufficiently representative of marginalized identities. Make your best efforts to allow for, and adjust to, feedback along the way.

We wish you the best of luck with your IA conference planning endeavours!

# Timeline

Organizing an IA conference is a difficult and time-consuming task, and the goal of this section is to provide a rough suggested timeline for future organizers to follow. This timeline was built from what was learned from the IA1 and IA2 meetings.

## 14–18 Months Before

- Identify members of the [organizing committees](#) (OCs)
- Understand how to address [accessibility](#) throughout the planning process
- Create an email list and online collaborative work directory
    - For communication within OCs
    - Be mindful about privacy policies of such an online file storage/file sharing application
- Determine the goals/themes of the conference (see [Programming](#))
    - Consider if there are upcoming events that would benefit from the output of the conference (e.g., Decadal Survey)
    - Consider if there are relevant current events that are impacting the astronomy community
    - Think about what actions and conference structures can produce the desired outcomes/deliverables that are in line with those goals
    - Brainstorm topics and presenters
    - Depending on the scope/goal of the meeting, choose topics and potential invited presenters
- Determine location and/or host organization (see [Spacetime Coordinates](#))
    - If your institution has an event planning team/group, then start working with them on a regular basis (see [Logistics/Events Planning](#))
    - Carefully consider number of participants and [accessibility concerns](#)
    - The location will need to be accessible and could potentially need a health room, quiet room, gender neutral/non-binary bathrooms, and child care facilities in addition to a large auditorium, poster display space, and multiple smaller conference rooms (if planning to have parallel sessions) based on the goals of your conference
    - Consider venue's A/V system for [remote participation](#)
    - Consider implications of any venue charges, if any, on your budget
- Determine rough dates (see [Spacetime Coordinates](#))
    - Consider impacts of school holidays, religious holidays, other major conferences (specifically those on related topics), and other events
- Start [budgeting the expenses](#)
    - How much will registration fees, lunches and breaks, conference dinner etc. cost?
    - Will there be lower registration fees for participants who need financial support?

- - How much is the home institute's overhead?
  - Any financial support for participants?
  - Is travel support going to be offered, if so, how and when will this be allocated and reimbursed?
- Look into funding opportunities from NSF, AURA, AAS, home institute, local businesses
  - A clear idea of the conference goals, logistics, and finances is needed for most funding applications

## 10–14 Months Before

- Design a logo and promotional materials
- Create website
  - Make sure it is accessible (on computers and mobile devices) and easy to navigate
  - Create an accessibility information webpage regarding the venue and resources that will be available
- Make a "save the date" announcement (including a conference flier)
- Think about where to advertise (see External Communication)
  - Email lists from past meetings
  - Different institutes, organizations, groups
  - Through AAS, AstroBetter, social media, etc.
- If considering a pre-registration process
  - Pre-registration should open when "save the date" announcements are distributed
  - A pre-registration process could give some idea of number and demographics of likely participants, especially if venue has limited seating (see Spacetime Coordinates), but it could also mean that you might have to do some kind of selection for conference participants or plan for a different venue
  - Plan and share the procedure you'll be using to select participants
- Finalize dates and venue

## 8–10 Months Before

- Plan logistics: hotels, catering conference breaks/lunches, conference dinner, social event (if any), transportation
- Work on anti-harassment policy, ground rules/group agreements, Slack code-of-conduct etc. to be shared on the conference website and during the conference

## 5–8 Months Before

- If using a pre-registration process, implement proper procedures for selecting participants

- - Consider all pros and cons of your selection process and if needed, consult with past meeting organizers, other community members/groups, or social science experts/professionals external to astronomy
  - If using [remote participation](#) (which may include remote presentations) during the conference, then start planning with the A/V team using proper guidelines or a participation agreement
  - [Poster presentation](#) details and planning
    - Consider how much space is available to display physical posters, whether to use electronic posters only or combination of both, lightning talks for poster presenters
    - Depending on space and electronic facilities, decide whether all participants can bring a poster or only selected ones
  - Finalize [logistics](#) (hotels, catering, conference dinner, social event, transportation)

## 3–4 Months Before

- [Select/finalize presentations and talks](#)
  - Confirm invited speakers
  - Inform other speakers and poster presenters
  - Email all presenters with proper information/guidelines (e.g., talk duration, Q&A format/duration, presentation format, accessibility requirements, permissions if presentations are to be made publicly available)
  - Inform other participants if their talks or poster abstracts are NOT selected
- Create and release a preliminary schedule for all participants to be able to assess whether they would like to register for the meeting
- Open (final) registration for all OR for [selected participants](#)
  - Give choice of in-person or [remote participation](#) (if offering)
- Make sure all communication channels are open (email, Slack, website, etc.) and the website has all details up-to-date (see [External Communication](#))
- Publicize the [travel](#) and dependent care support deadline
- Arrange to have trained people to respond to harassment complaints during the conference
  - If possible, designate people (e.g., from HR) who could be the points of contact if any issues arise during the meeting
- If the program includes discussion sessions, consider arranging for trained moderators and/or social scientists to aid with constructive discussion
- Put together codes of conduct for the conference and ground rules for in-person and online discussion, make these available on the conference website

## 2–3 Months Before

- Close [registration](#)
- Check on [budget](#) with the registration closed

- - Depending on how many participants registered, you could think about late registration, for additional participants/presenters, or targeted invitations, inviting specific individuals or groups
  - If hiring a photographer, finalize contract

## 1–2 Months Before

- Finalize the [program](#) (for any last minute changes)
  - Get confirmation from all presenters, and if needed (e.g., because of last minute declines), send additional speaker invitations
  - Confirm/finalize details about [remote participation](#) and/or presentations
- Finalize all [catering needs](#), e.g., breaks, lunches, conference dinner
  - Make sure locations of these meal stations/venues meet [accessibility requirements](#)
  - Make sure all dietary restrictions are taken into account
- Send participants any specific or additional [travel support](#) and dependent care details

## < 1 Month Before

- Assemble and distribute conference [welcome packets](#)
- Generate Google maps (or other resource) with local places tagged (e.g., places to visit, restaurants, bars, activities)
- Prepare to make last minute changes to the program because of participant's schedule changes e.g., cancellations or switching from in-person to [remote participation](#), especially if that involves [travel support](#)
- Allow for people to update their registration information as it relates to badging, i.e preferred name, pronouns, meal restrictions, anything that might have changed since the initial registration
- Plan an [exit survey](#), to be distributed to participants during the conference or immediately after the conference

## 1–2 Weeks Before

- Inform speakers and poster presenters where to upload talks and electronic posters
- Check all presentations (talks, posters) on A/V systems
- Test presentations and connections for [remote participants](#)
- Prepare/print various signs: registration desk, rooms, bathrooms, security doors/exits, wheelchair accessible entry/exits, maps, etc

## 1–3 Days Before

- Dry run of any necessary logistics and discussion of who is responsible for what

- Consider creating a schedule for LOC members to be assigned to various roles throughout the meeting (e.g., monitoring Slack, overseeing remote participants, passing microphones)

## During the Conference

(see [Running the Conference](#))

## 1 Week After

- Ensure participants' expenses are reimbursed for any [financial support](#) they were allocated (if travel advances were unable to be issued)
- OC check-in on how the meeting was: conference organization, structure, program, talks, posters, logistics, breaks, interactions/discussions, etc.
- Send email to attendees: thank them for participating, ask them to fill out survey
- Rest and recuperate!

## 2–4 Weeks After

- Review exit survey results and responses
- Create a plan for summarizing various aspects of the conference, e.g., community recommendations, future meeting recommendations

## 1–6 Months After

- Write summary documents
  - Recommendations for conference planning, please feel free to use this document as a base and continue to evolve these recommendations
  - Conference outcomes and community recommendations
- Have documents reviewed by external people
- Send thank you notes to conference sponsors

## 6 Months After

- Publish and distribute summary documents

## 6–18 Months After

- Write any final reports for the institute/grant agencies

# Organizing Committee: Structure, Roles, and Responsibilities

In order to make an IA conference as inclusive as possible, the organizers will need to give their attention to a wide range of logistics. Traditional organizing committees (OCs) for conference planning (e.g., a Science Organizing Committee and a Local Organizing Committee) are typically not structured to address all of the needs of marginalized identities. To ensure these needs are met, we recommend a single OC that oversees all aspects of conference planning. In this section, we provide both general recommendations and an example set of "focus areas" for subcommittees and committee members.

## Membership

- The OC should be as diverse as possible, with representation across marginalized identities, expertise, geography, etc.
- We strongly suggest reaching out to the [American Astronomical Society committees and working groups](#) that work directly on inclusivity in astronomy (e.g., WGAD, SGMA, CSMA, CSWA). These groups will be able to provide valuable input on various aspects of planning IA conferences, they may wish to be part of (or consulted on) the planning process, and they could be a source of (or be able to suggest) possible committee members to ensure adequate representation on the OCs.
- Include members with experience in organizing inclusive meetings (not necessarily a previous IA).
- We suggest including at least one person with professional expertise in social sciences.
- Carefully consider how many people are on the committee, making sure that the total number of people stays at a manageable size. Otherwise, tasks can inadvertently get dropped and/or a small group of people end up taking a disproportionate share of the work.
- Include at least one member from the previous IA OC(s), either as a full member or in a consulting role, to prevent loss of institutional memory.
- Consider a diverse range of ages, experience, and career stages within the OC and how the membership should balance the need for established DEI leaders as well as mentorship of the next generation of leaders.

## Committee Structure

- We suggest two tiers of leadership and area focused subcommittees:
    - Chair: 1–2 people, responsible for leading the committee, ensuring workload is evenly distributed, and overseeing the high-level goals of the conference are met.

- - Co-Chairs: Several people, each responsible for 1-2 broad focus areas (listed below). Each focus area should have several committee members working on the tasks in subcommittees.
  - Subcommittees: Groups of people led by co-chairs to focus on broad areas of organization. Possible focus areas could resemble goals of traditional local and science OCs, e.g., dedicated subcommittee for setting up the conference program and subcommittee(s) for conference planning logistics.
- OC roles can be in addition to what is provided by events organizers at the [venue](). However, constant communication between the events teams and the OCs, and an understanding of what the responsibilities of each are, is very important to avoid oversights.

## Focus Areas

We recommend that each of the broad areas listed below have at least two organizing committee members assigned to it (including a co-chair) to avoid single-point failures. Links to their relevant sections are included for more information on what each role could involve.

- Conference content
  - [Setting the conference programming]()
  - [Communication with contributors]()
- Participant logistics
  - [Registration]()
  - [Remote Participation]()
  - [Communication with participants]()
- Facilities planning
  - [Accessibility]()
  - [Event Planning]()
- Finances/Money
  - [Expenses]()
  - [Funding]()
- Social contact(s)
  - [Social media and communications]() (Slack, Twitter, Facebook, Instagram, hashtag, LinkedIn, website, etc.)
  - [Social event]() (i.e. conference dinner) planning
- During conference
  - See [Running the Conference]()

# Accessibility

The best and easiest way to approach accessibility is to include and design for it in every part of the planning process.

Many times well-intentioned decisions are made that must later be retrofitted into something more accessible to neurodivergent people, people with disabilities, and the Deaf and Hard of Hearing community, usually with a greater expense of time and effort than would otherwise be needed. From the very earliest decisions, such as when to schedule the conference and what the venue should be, to the last minute decisions like making and putting up directional signs, being cognizant of accessibility concerns in your decision-making process will lead to a conference which better serves the needs of the community.

IA2 ultimately found that an effective combination was having a single individual from the LOC tasked with overseeing and remembering accommodations during all parts of organization and planning, as well as consulting an expert in the field of accessibility from the [AAS Working Group on Accessibility and Disability (WGAD)](). While it's not necessary to adopt this setup specifically, we do recommend that you formalize a method to consistently evaluate ideas and decisions for accessibility.

When adopting this holistic approach to accessibility, remember to think broadly about designing accommodations. If only one type of accommodation is considered, such as wheelchair accessibility, then a significant number of attendees will not have the other accommodations they need, because having ramps will not address a Deaf attendee's need for captions. The goal is to make all parts of the conference as accessible to as many people as possible.

Here are some resources and practical tips as a starting place for accommodations and accessibility:
- The WGAD conference checklist contains prioritized lists of action items and a timeline. [Contact WGAD]() members for the latest version of this document. Please consider this essential reading.
- [The IA2 Accessibility webpage]() has additional specific guidance and resources on meeting accessibility.
- [The IA2 Accessible Presentation Checklist]() for information on making and giving accessible presentations.
- Be transparent about what you can and cannot provide. Never promise an accommodation that isn't researched and confirmed.
- Give out information about the accommodations as early as possible to help those who have accessibility concerns advocate for themselves and make plans.

- The [additional resources section](#) has links to more information on improving accessibility at conferences from other organizations.

# Programming

This section lists recommendations on how to select contributions and decide the content for the conference. While future IA organizers will come up with their own criteria, the recommendations below may help support this critical phase in conference planning.

## Considerations for Programming

- Widely distribute a community survey to see what people expect and want from the next conference. This will inform topics, balance of talks/discussions, venue size, etc.
- Decide early on what the specific goals of the conference are, and have the programming choices follow from that.
- Talks should be selected and confirmed well in advance of the registration deadline registration for other participants so that potential participants have the schedule before making a decision.
- In addition to talks, having panel discussions among speakers or invited panelists can work well. Allow sufficient time for questions and discussion and keep intersectionality in mind when selecting panel members.
- Include some interactive content (e.g., unconference sessions, subject-specific moderated discussions, hack days, networking). This was highlighted in the IA2 post-conference survey as something that was of particular importance to participants and that we were unable to address sufficiently with IA2.
- Ensure all breaks are at least 30 minutes long and lunches are at least 90 minutes long.
- There should be an auditorium or open space that can seat the total of all attendees for talks. There could also be rooms for concurrent talks and workshops held in parallel (see [Determining Spacetime Coordinates](#)).
- If a talk is on determining paths forward, be sure to ask the presenter to give specific recommendations in their talks, so they can provide the basis of discussions and/or conference recommendations.
- Understand which speakers will participate remotely (see [Remote Participation](#)) and what their needs are. To ensure equity in conference participation, a speaker should not be disqualified if their participation is remote.
- When finalizing the conference schedule, put remote presenters immediately after breaks to ensure final checks of their setup prior to their talk to avoid delays. Communicate with remote presenters to factor in their timezone and scheduling constraints.

## Selecting Contributions

- Set criteria and guidelines for selecting speakers and posters, and make them public *before* the request for speakers. Allow feedback, and listen to the community if these criteria could be made more inclusive or improved if deemed not inclusive.
- For invited talks, the OC should do their homework and consider speakers that could talk on subjects that align well with the conference goals, are respected within the community as an authority on the subject, and that are verified to be able to address the subject conscientiously.
- Have a clear way for people to indicate that they want to give combined/complementary talks (perhaps in the [Registration](#) form).
- In addition to those in our field, be sure to invite speakers from other fields to give talks (e.g., race and gender studies departments). This can be challenging but very beneficial.
- The IA2 SOC selected talks using the dual-anonymous procedure currently in place for Hubble Space Telescope proposals. Unfortunately, this resulted in speakers with privileged identities being over-represented, and in some cases, speakers presenting about identities they do not share. There are significant benefits to selecting content without the typical unconscious biases (such as the balance between new and established voices), so we suggest a modified procedure for future IA conferences. We propose that in addition to submitting abstracts, there should be a space for potential speakers to explain why they are qualified to present their talk. For instance, if the talk is about being inclusive of the disabled community and the speaker is not disabled, they should explain why they, and not a disabled individual, should present material about that community. The OC would need a procedure to ensure confidentiality, particularly for invisible or hidden identities. After anonymized abstracts have been coarsely ranked, this additional information can be used to finalize the program by ensuring good representation and balance of identities.
- This talk selection procedure will likely play into your choice of whether to collect demographic data (see [Demographic Information Collection](#)).

# Determining the Spacetime Coordinates

This section will cover our recommendations on choosing the location and venue for future IA conferences and considerations for conference scheduling.

Note: The spacing recommendations are assuming that the conference will be in person, but with the coronavirus pandemic, we note that a number of conferences are moving to be entirely online. While we hope that future IA conferences will still be possible in-person, at least in part, we recommend viewing the section on [Remote Participation](#) that contains some relevant advice for fully remote meetings. In light of the global pandemic, we certainly recommend that organizers make contingency plans for moving future meetings to be entirely online as required.

## Spacing

- **All spaces should be accessible** (see [Accessibility](#)).
- IA1 in 2015 had 150 participants, and members of the IA1 OC mentioned that it was a challenge to reach those numbers. IA2 had about 300 people sign up for pre-registration, while only having conference space for 180 people. We encourage future organizers to think carefully about the scale of their conference and perhaps delay choosing a venue until the interest level can be gauged.
- A decision on the capacity of the venue is something that should be carefully considered. The interest in the conference series appears to be growing quickly (which is exciting!), so time should be spent evaluating how to best include the most voices based on the goals and structure of the conference. A smaller conference could allow for more focused conversations, but can also limit the breadth of those conversations.
- We recommend that when selecting a space you keep in mind that the number of people attending the conference will dictate or be dictated by the following (this will impact [Registration](#)):
    - Do you want all participants to attend all talks/sessions (see [Programming](#))?
    - How can you ensure meaningful discussions occur that include all participants?
    - Parallel sessions can expand the range of topics that the conference addresses, and smaller groups may aid discussion, however, some people that are interested in, or can benefit from, sessions held simultaneously may be excluded.
    - How can you ensure the room and conversation is not dominated by more privileged groups?
- Availability of social, outdoor, and quiet spaces are highly recommended.
    - Socialization among attendees helps foster community and allows open discussion of the conference content. Attendees of IA2 gave feedback that they desired more time to digest and discuss the various topics with their colleagues, and emphasized that networking was potentially the most important component of the conference. We recommend a large social space, in accordance with [accessibility](#) guidelines (e.g., not too echoey).

- - The health and wellness of the attendees can be fostered by having outdoor spaces and quiet spaces.
  - A quiet space should have clearly posted guidelines for use (e.g., no talking, no phone calls) and could contain resources for quiet activities.
- There may be attendees that will need to bring their children. Making efforts to have a space for children and caregivers is recommended.
- Have a comfortable, private space for lactating (nursing or pumping) mothers.
  - Mothers with nursing children ideally would be able to nurse anywhere (as the venue allows), but having a private space is an important option.
  - There are many considerations when creating a lactation space. A good example of best practices can be found [here](#).
- Storage space for luggage should be available from the start to end of the conference.
- Consider venues with all gender bathrooms. If all gender bathrooms are not available, work with the venues to mark some bathrooms as all gender for the duration of the conference and create the appropriate signage for the bathrooms.
- Accessibility should be considered for transportation, including the distances between venues and hotels. Wheelchair accessible transportation should be explored.

## Timing

- At minimum, the conference should be 3 days long. IA2 was 2 days long and the *overwhelming* feedback we received was that it needed to be longer.
- Schedule the conference keeping in mind potential conflicts with holidays and other conferences in the astronomical community.
- Plan time for social interactions. This allows the community to engage with each other with the topics of the conference, fresh in mind.
- Timing of breaks and lunches should be strategic, regular, and at least 30 minutes for breaks and 90 minutes for lunch.
- Depending on distances between conference locations, allow appropriate time between sessions and events for attendees to travel. Pay particular attention to timing when traveling between venues for attendees with accessibility needs.

# Logistics/Events Planning

In planning an IA conference, there are many decisions and considerations at every stage. This section outlines the things that the organizers will need to address, from the earliest stages of planning, to activities during the conference, through to wrapping up in the weeks and months afterwards. We hope that this section, which contains a combination of broad considerations and narrow minutiae, will help both the planning and the conference itself proceed smoothly.

## Broad considerations before starting

- For [corporate-funded](#) large events, you need ~1.5 years lead time.
- You should begin planning at least a year before the conference (see [Timeline](#)). Less time than this will make it difficult (and in some cases, impossible) to accomplish all of your inclusivity goals.
- There are other conferences related to DEI (Diversity, Equity, and Inclusion) that you could draw from when designing future IA meetings, e.g., [Women in Astronomy](#), [Tech Inclusion](#), and more.
- AAS isn't able to organize one of these conferences due to available resources, but if one is attached to a regular AAS meeting, they might be able to help with logistics (secure a space, organize catering, etc., as was done for past Women in Astronomy meetings).

## Setting up planning structures

- Choose a platform for taking notes at OC meetings, collaborating on documents, etc.
- Have regular OC telecons using an accessible platform. These may be every two weeks in the earliest stages of planning, but should be weekly starting ~6 months before the conference. In the last 3-4 weeks, you may need to meet twice weekly.
- For additional details, see the [Organizing Committee](#) section.

## Event planning

- If your institution/chosen venue has an events planning team, be sure to communicate with them early and often. You may be implementing many new options that the events team is less familiar with (e.g., pronouns on name tags, different tiers of registration fees), and they will need time to accommodate these changes. Also, be clear about what is the event team's responsibility and what is the OC's responsibility.
- If you do not have an events team (and in that case, we strongly encourage you to hire a professional planner), here are some things to consider. With each of these, be sure to keep [accessibility](#) in mind.
    - Procedure for collecting registration information and fees.

- - Identify hotel(s) for participants to stay in, and arrange discounted group rates.
  - If you're hosting at a university, depending on the time of year, you may be able to utilize low-cost housing in dormitories.
  - Arrange for on-demand accessible transportation between hotel(s), conference venue, social outings, and/or conference dinner. Ensure there is an adequate number of shuttles to transport everyone in a timely manner.
  - Provide information about how to get to the conference location and venue.
  - Ensure organization regulations and codes of conduct are shared with (and where necessary signed by) participants.
  - Communicate information about booking, travel funding support, and any limitations (see Funding and Expenses sections).
  - Formatting and printing name badges.
- If the conference is at a convention center
  - There are people whose job it is to hang signs, arrange catering, etc. Take advantage of their expertise and use your time/energy to focus on other things.
  - Keep in mind that corporate entities may have challenges with last-minute changes. This is particularly relevant if you won't know the needed accessibility accommodations until the participant list is finalized.
- Decide if you would like to have an optional pre- or post-conference social outing. This is a good opportunity for networking and to explore the conference city/region (e.g., including justice-related themes if applicable). When planning, consider accessibility (both the location and transportation) and cost. Communicate the timing to attendees early enough so they can consider participation before buying plane tickets.
- Consider hiring or sourcing a professional photographer for a conference group photo and mid-conference photos.
- The success of some logistics are predicated on participants being aware of them. Think carefully about how you will communicate these to participants, since people are unlikely to read long explanatory emails (see External Communication).

## Food and meals

- Ask attendees about dietary restrictions during registration, and use that information to ensure that all meals/breaks have food acceptable to everyone. Note that these may change between registration and the conference, so allow for participants to update their requirements.
- If an attendee has a severe allergy, it may be necessary to eliminate that item from all meals and snacks (including those brought by other participants).
- When selecting a caterer (assuming your venue allows a choice), consider choosing businesses operated by people with marginalized identities.
- Determine a method to efficiently distribute food so that people aren't required to stand in line for extended periods.
- Ensure all food is labeled with ingredients and possible allergens.
- During meals/breaks, consider setting up discussion tables:

- - Folks who want to meet new people and/or are attending alone
  - Topic-specific tables
  - Optional identity-specific tables
- Have water and shelf-stable snacks (e.g., pretzels, nuts, fruit, cookies) available to easily grab at any time during the conference.

## Running the conference

- At the start of the conference, take 10-15 minutes to go over logistics
  - Outline the goals of the conference
  - Go over ground rules, codes of conduct, discussion guidelines (examples for IA2 [here](#))
  - Territorial and labor acknowledgements
  - Locations of bathrooms, quiet space, breaks, meals, posters
  - Points of contact for different things: accessibility, code of conduct violations, general logistics
  - Explain how [remote participants](#) will be engaging in the conference
- Logistics for talks
  - Have someone responsible for getting talks transferred onto the main presentation computer (recommended) and/or testing speakers' own computers.
  - Duties for session moderators
    - introduce presenter (ensure correct name pronunciation and pronouns)
    - make sure presenter is using a microphone, and ask them to adjust microphone mid-talk as needed
    - monitor the time
    - lead Q&A
  - More advice on chairing meetings is available in the [additional resources section](#).
  - Depending on the size of the room, have 2–4 people carrying microphones to the people asking questions. Have backup microphones/batteries available.
  - Have someone monitoring for online questions (either from remote participants or from participants who prefer not to ask questions aloud).
  - For the speaker, consider using an over-the-ear microphone, rather than one that clips on clothing. An ear microphone stays at a constant distance from the presenter's mouth, so it is much clearer, especially for people online or those in the Hard of Hearing community.
  - Have one or more people designated to provide technical support: computer issues (PowerPoint/Keynote), A/V (projector, sound system), teleconferencing software (for remote participants).
- If there are remote participants, someone will need to be available for the entirety of the conference to communicate with them about any connectivity or other issues that arise (see [Remote Participation](#)).
- There will likely be instances of presenters using non-inclusive language. Assign someone, preferably more established in the field, to be responsible for responding.

- Have someone available (perhaps an HR representative from your institution or external professional) to address code of conduct violations or other serious issues.
- Immediately before and during the conference, have 1–2 OC members monitoring Slack discussions for problems and/or questions (can rotate duties). Consider involving people who are trained in human communication (e.g., social scientists, professional discussion moderators).
- During breaks, have 1–2 OC members as a designated "point person." They will handle any questions or issues that arise. The OC members not on duty during the break can then be free to enjoy the conference (socialize, look at posters, etc.).
- If needed, have someone to open/close doors between talks to facilitate getting water, going to the quiet room, ensure fresh air flows through, etc.
- Make sure people define their terms/acronyms. Not everyone knows what terms like R1/R2 colleges (USA terminology) and DEI (diversity, equity, and inclusion) mean.
- If discussions are not being recorded, have somebody assigned to take notes, so that discussion points can be incorporated into conference recommendations.

## Wrapping up after the conference

- Send out a [post-conference survey](#) for feedback.
- Within the organizing committee, discuss lessons learned and things that did and did not work at the meeting. Write them down as soon as possible so you can remember them for later.
- Build upon and update the recommendations in this document.
- Distribute these recommendations to potential future IA organizers. Also consider distributing more widely in the astronomy community.
- See also the [Timeline](#) section.

# Expenses

In this section, we will cover expenses that are expected for any IA conference. This section is only to provide a general guideline of what to consider and does not cover all possible expenses. We recommend that more than one person be tasked with budget oversight. We also would like to place an emphasis on expenses related to accessibility, dependent care, and early career support. Prioritizing expenditures on these will help remove significant barriers to an individual's ability to participate in the conference. As always, make contingency plans for unexpected and smaller expenses.

## Registration Fees

IA2 utilized a two tiered registration fee approach:
- Early career participants paid a reduced registration fee to support goals of the conference. We recommend making this optional as some early career individuals may be able to and volunteer to pay the full registration.
- Mid and late career paid full the registration fee, and this was used to offset the cost of early career registration.
- We recommend developing a process to allow anyone to request reduced registration fees.

## Transportation

- Consider what resources are needed to increase [accessibility](), e.g. wheelchair accessible shuttles.
- We highly recommend that any booked transportation includes wheelchair accessibility. Consider if the selected venue offers accessible transport and if not, consider this in the budget creation.

## Venue

- We highly recommend a cross-check of [pre-registration]() numbers (if using) with venue capacity.
- Improvements may be needed to increase accessibility.
- Percent overhead cost: Depending on the venue and also where funds are dispersed, there are certain overhead expenses that can be incurred. For example, if the AAS helps to fund the conference and money is deposited into a local account (e.g., at your home institution), then a standard overhead expense may be deducted from the AAS funded amount.
- Extra expenses can be incurred outside of space rental (e.g. furniture rental).
- Printed materials

- Name badges
- Printed programs
- Signage

## Food

- Gather information from pre-registration to select a food vendor that meets the needs of all participants (see [Logistics/Events Planning](#)).
- Discuss expenses with vendors once identified.
- Certain venues require the vendors to bid if expenses are greater than a certain amount.
- Account for dietary requirement options which may increase catering costs.

## Lodging

- Provide participants easily accessible shuttles to travel back-and-forth between the hotel and event locations on demand (see [Event Planning](#)). Account for this travel frequency when selecting the form of transportation and hotel location (e.g., more distant hotels may require more and/or higher-capacity shuttles).
- Consider whether there will be multiple "official" hotels. These can accommodate a variety of participant budgets and needs, and less expensive options can stretch your travel support budget. This could also increase the cost of any conference-provided shuttle transportation.

## Travel Support

- One goal of both IA1 and IA2 has been to allocate travel support to early-career individuals. For IA2, we provided a reduced registration fee and funding for travel support to encourage early-career participants to attend the conference.
- Consider how many participants will receive support, with an emphasis on supporting individuals who typically would not be able to attend a conference.
- Financial assistance for dependent care: Review terms and conditions from funding sources to see if this is an allowable expense (note that dependent care is not allowable for NASA/NSF funding sources). If this is not an allowable expense, consider offering resources of local care providers and allow [remote participation](#).
- Not all participants can afford to wait weeks or months for their expenses to be reimbursed. Either ensure that reimbursements are disbursed rapidly, directly pay for travel costs, or offer travel advances.

### Determine the number of participants who qualify for travel support

- If you are using a [pre-registration](#) form, you could collect information there about any travel support needs.

- It is up to the OC to determine the selection criteria for qualification (See [Participant Selection](#)). Qualified recipients could be, e.g., early career, anyone who requires assistance, etc.

### Transparency

- Send acceptance letters for travel support early once qualified candidates are identified.
- Make sure all conditions for the funding are known upfront prior to any reservations being made to avoid people losing money on expenses that don't qualify for reimbursement (e.g., some funding sources require American-based airlines).

### Amounts

You could choose to have a funding cap per individual or make it variable.
- **Fixed:** As adopted in IA2 (partial support $500, full support $1000).
    - Pros: We were able to determine the maximum amount that could be allocated from the budget. When a participant declined, we knew the exact surplus which could be allocated to another participant for partial or full support.
    - Cons: When conference participants submitted their travel expense reports, the exact dollar amount was either under or over the allocated amount. This puts funding at risk of under- or over-utilizing allocated funds for travel support. If there is a surplus, this could have been used to increase support for more participants.
- **Variable:** This option was considered but not used for IA2.
    - Pros: Naturally accommodates uncertain expenses, e.g., flights changing price.
    - Cons: There would be a risk involved with the account becoming negative. Contingency would need to be considered to prevent accounts from being in a deficit. There are many logistical constraints with this option as it requires up-to-date booking information from all funded participants at all times.

Determine a process for reallocating travel funds when a participant declines. Two to three months before the conference, close conference registration and finalize all travel support. If anyone declines after registration closes, reallocation can become difficult and run the risk of under utilizing funds. It is at the discretion of the organizing committee to allocate appropriate funds for any travel support; however, we recommend continued support for early-career individuals and dependent care where possible.

## Budget Timelines and Considerations

- [14-18 months](#) before:
    - Start the budgeting process and identify large and important expenses.
        - [Venue](#)
        - [Food](#)
        - [Travel Support](#)

- ○ The conference registration can be used as a funding source to offset the fees for early-career participants.
- ○ Factor in the number of participants after pre-registration closes to obtain a budget upper limit. This number will be adjusted periodically throughout planning (see [Participant Selection](#)).
- ○ Determine how funds will be allocated and dispersed. Certain funding sources require an invoice reflective of the original agreement. (e.g., if you requested funds to support early-career participants, they might disallow spending funds on participants at other career stages).
- ○ Periodically update the budget and stay in close contact with necessary parties.
- [2-3 Months Before](#):
  - ○ Check on budget with registration closed.
  - ○ Collect all invoices that will be needed for itemized expenses covered by sponsors
- [6-18 Months After](#):
  - ○ Submit final report on expenditures.

# Funding

In this section, we will cover recommendations for seeking funding. It is important to establish conference details before reaching out to any source for funding. It is also beneficial to create a preliminary budget to have an idea where funds will be needed the most. One important resource limitation is space (see [Spacetime Coordinates](#)), so we recommend accounting for flexible venue options in funding, if necessary.

## Sponsors

Identify and contact potential sponsors. Possible criteria include:
- Groups aligned with the goals of diversity, equity, and inclusion
- Organizations, companies, or entities that are astronomy-related
- Companies that could hire astronomers leaving academia (especially ones in the location where the conference is happening)
- Childcare: We highly recommend the committee determine if childcare is an allowable expense from any funding source (see [Travel Support](#)).

## Funding Timelines and Considerations

- [14-18 months](#) before: Start writing any proposals (e.g., National Science Foundation)
    - Include number of attendees from previous IA conferences
    - Present the outcomes from previous IA conferences
    - If needed, disclose other sources of funding
- Sponsors may require proof that the conference is occurring (e.g., any forms of advertisement, save the date, etc.) before providing any funding.
- Be aware of any contracts in regards to conference cancellations.
- Show gratitude to sponsors
    - Display sponsors' logos on website, in announcements, and at the conference
    - Send thank you letters after the conference

# Remote Participation

To ensure your IA conference is as inclusive as possible, consider implementing remote participation if your setup and venue can support it. Remote participation has been shown to increase diversity in conference participants ([Prichard et al., 2019](#)). There are multiple levels of remote participation and you may consider implementing one or both of the following: 1) a one-way public webcast for talks to be streamed (webcasting service, e.g. Panopto, YouTube), 2) a two-way remote participation link for conference participants unable to attend in person (video conference software, e.g. Webex, Zoom). A full report on the usage of remote participation at IA2 is available [here](#), below we summarise the key recommendations and a step-by-step guide for its implementation.

Note: The coronavirus pandemic has resulted in an increasing number of conferences moving entirely online. While the following recommendations are targeted towards hosting a combined in-person and remote participant conference, we envisage much of the advice below will still be relevant for fully remote meetings. For more specific advice on hosting all online meetings, the [Committee On INclusion in SDSS (COINS)](#) has some [recommendations for hosting and moderating telecons](#) that also has some relevant advice (also see [Additional Resources](#)). We anticipate more recommendations for purely remote meetings will arise from the conferences occuring online this year.

## Recommendations for Remote Participation

Based on remote participation usage at IA2, the following are high-level recommendations for its implementation at future meetings.

- Consider having two modes of remote participation:
    - one-way public webcasting where talks are streamed live and recorded,
    - two-way interactive remote participation for conference participants unable to attend in person.

### One-way public webcasting recommendations

- Consider recording talks. Recordings allow people who could not attend in person or participate online during the conference to access the information later.
    - During IA2 no one followed along with the livestream but as of November 19, 2019, ~1 month after, there were 541 views or downloads of all the available videos.
- Ensure you have express permission from all presenters for their talks to be recorded and publicly webcast, and allow this to change at any time. Presenters may opt out of sharing/recording their talks for a number of reasons including sensitive/personal content or proprietary information.

- At IA2, 2 out of 24 presenters opted for their talks not to be publicly webcast.

### Two-way remote participation recommendations

- Ensure you have express permissions from all presenters for their talks to be shared with the two-way remote participants.
  - At IA2, all presenters opted for their talks to be shared with the two-way remote participants.
- Allow for in-room participants to switch to two-way remote participants up to and during the conference.
  - Prior to IA2, 12 people were registered as two-way remote participants and 4 local people indicated they may join both in person and remotely.
  - During the meeting, a further 3 people requested to be two-way remote participants due to last-minute changes in personal circumstances.
  - Of these 19 people expected online, 13 people connected over the course of the conference.
- Give the option for presentations to be given remotely.
  - There were 3 remote presenters at IA2.

From the [post-conference survey](#), the overall level of satisfaction with remote participation was 7.8/10.

## Setting Up Remote Participation for Meetings

Given the relative success of running remote participation for IA2, and learning from the feedback we received, below are a suggested set of steps for organizers on the ways to implement it for future conferences.

### 14–18 Months Before

1. Set up a meeting with the audio-video (AV) team at the chosen venue at the start of planning the meeting to discuss the possibility of remote participation.
2. Find out the AV setup for the space and remote participation options available far in advance of the conference. If the AV setup is insufficient in any way (e.g., bandwidth, software not set up, hardware not available), investigate ways it could be implemented within the time and budget constraints.
3. Explore [accessibility](#) options for remote participation such as live closed captioning, applying captions to recorded videos in post processing, adequate sound quality etc.
4. Set a price point for remote participation registration, either:
   a. free,
   b. at a low level to deter large volumes of people opting for remote participation (if needed),

   c. at a level to cover any costs involved in implementing remote participation (if they exist).

## 10–14 Months Before

5. Once the arrangements are set, and preferably prior to registration, offer two-way remote participation as an option for all invited participants and/or offer the option of the public webcast as a one-way remote participation option for everyone.
6. Post information about remote participation options on the conference website along with any relevant links and distribute further if required (example for IA2 [here](#)).

## 3–4 Months Before

7. If using, ensure that all two-way remote participants have signed/acknowledged a remote participation agreement (example for IA2 [here](#)).
8. For presenters, ensure they have completed their forms granting permission to be recorded and publicly webcast. For any presenters not wishing to be publicly webcast, ask if they agree to their talk being shared with the two-way remote participants (and send them a copy of the agreement for reassurance if required/using).

## 1–2 Months Before

9. When finalizing the conference schedule, put remote presenters after breaks to ensure final checks of their setup prior to their talk to avoid delays.
10. Be mindful of the remote presenter's and participant's timezones when setting the schedule.

## < 1 Month Before

11. Around 3 weeks before the meeting, meet with AV to discuss the final arrangements of the remote participants, how many two-way remote participants are expected, how many presenters are remote, if and when the public stream should be switched off etc.
12. A few weeks in advance, set a time the week before the conference for all remote presenters to run checks on their video feed, presentation, audio connection etc. with the AV team in the conference location.

## 1–2 Weeks Before

13. Two weeks in advance, get AV to set up links for the one-way public webcast and two-way remote participation. Post the links for the public webcast on the conference website and send to anyone interested in following along via the one-way webcast link.
14. The week before the conference, share the two-way remote participation link with all two-way remote participants along with some basic information, conference schedule, and tips for connecting/presenting remotely that can be crafted with AV. Example of these tips is below:

      a. Always use a headset/microphone to ensure the best sound quality
      b. Be aware of your environment
      c. No backlight
      d. Minimize background noise
      e. Mute your mic unless talking
      f. If possible use a wired network
      g. Be aware that you are likely to be seen by many other people including those attending in the auditorium.

15. Create a live document of the schedule indicating when the public stream should be switched on and off (depending on the permissions from the speakers which may come in late). Send this to the AV team.

## 1–3 Days Before

16. Have a method for two-way remote participants to ask questions, either through the Webex (or other platform) monitor, or a more public forum (e.g., conference Slack, highly encouraged to increase connectivity between all participants) and ensure an organizer is monitoring this at all times.
17. If required, display the online participants, post their names to a conference workspace (e.g. Slack), or let other participants know who is in person and who is remote at the meeting (e.g. a participant list) to facilitate interaction.

## During the Conference

18. Connect to the two-way remote participation link at least an hour before the meeting starts to allow people to troubleshoot connecting to the system before the talks begin.
19. Ensure that the people online are only those expected as two-way remote participants.
20. Announce clearly at the start of the conference what the arrangements for remote participation are to all in-room participants.
21. Ensure all speakers are using a microphone, that it is switched on, and that the remote participants can hear the sound feed clearly. An over the ear mic has the best sound quality and consistency, then a lapel mic, and lastly a hand-held mic (only use this option if no others are available). Also ensure live closed captioning is enabled if available.
22. For all questions asked by in-room participants, ensure they have a microphone before talking or the question is repeated by the presenter/moderator into a microphone. For all questions submitted online by remote participants, the moderator should relay these questions to the room over a microphone.
23. During the meeting, indicate to the AV team when the talks should be publicly webcast and when they shouldn't. Even though they should have a schedule, this is a fail safe and also lets all the people in the room know when they are being streamed publicly. You may also wish to turn the stream off for questions to allow people to speak freely.
24. Monitor whether everyone online can hear the presenters and watch out for any issues they may be experiencing over the connection that could be fixed in the room.

## After the Conference

25. Ensure all recorded talks with permissions from presenters are available online (with closed captioning offered if possible) and publicize these with participants and more broadly if desired. Ensure the edited videos contain only those parts of the conference intended to be public (e.g. if deciding not to record Q&A).
26. After the conference, gather feedback from in-room and remote participants about their experiences at the conference, including remote participation, and look for ways that remote presentation could be improved.
27. Build on these recommended steps and pass onto future meeting organizers.

# External Communication and Materials

Communication and information sharing is important for all conferences but it is much more crucial for IA conferences. Here we share our thoughts on various avenues for external communication and what type of documents/materials might be needed.

## External Communication

Different people prefer different modes of communication, so consider making information available through multiple avenues.

### Website

- The conference website should contain all conference related information for the whole community, consider looking at past meeting websites for possible sections/categories for the web page.
- Decide beforehand what is to be shared and what is not, particularly regarding the program, talks/posters and names of the participants, ensure proper permissions are obtained.
- Make sure the pages are accessible and easy to follow.
- Allow 2–3 [OC members](#) to edit this page.
- Make sure pages are up-to-date.
- When appropriate, all information communicated by email and Slack should also be posted to the website.

### Slack/chat forum for participants/social media

- To promote a safe conversation space, set up a discussion space for registered participants only.
- Invite people to join after they have registered.
- Ensure a code of conduct for communication is prepared and posted (examples from IA2 [here](#)).
- A chat forum can be a place to chat and have discussions before/during/after the meeting.
- Possible channels
  - Place where all talks and posters will be shared (provided authors have given their permission to share)
  - Questions for OC members
  - Private channel(s) for discussion among OC members
  - Place for [remote participants](#) to ask questions, discuss topics, etc., (related to remote participation)
  - Place for people to introduce themselves

- - Make some topic-specific channels to get conversations started, and during the conference encourage participants to add additional topic-specific channels
  - Have moderators to ensure the code of conduct is being followed. LOC members can take turns to monitor the forum (especially important to allocate shifts during the conference)
  - If your institution has a social media liaison, consider having them regularly post information about the conference

### Email

- Have a conference-specific email address (not someone's personal email) to receive and send communication to individuals as well as to send general announcements.
- OC should members monitor this email account regularly.
  - Consider assigning 1–2 people to keep an eye on emails and redirect them as needed and rotate duties (if needed).
- When sending emails:
  - release information in as few emails as possible,
  - make sure the same information in emails is available on the website,
  - for long emails (which people are unlikely to actually read), consider including an outline or summary of the information.
- Have a system to manage email lists (e.g., people who signed up for notifications, people attending, people giving talks, people doing posters, people with [travel support](#), people with dependent care support)

### Online repository

(e.g., Google drive, Box, Dropbox) to share materials
- Conference related documents, maps, welcome packs, and other printed materials
- Talks and posters (after getting permission from the authors)
- Always be aware of privacy issues with online storage

## Materials

### Before the Conference

- Logo and style guide for the conference materials
  - This could be done with a graphic designer if resources or expenses allow
- 'Save the date' poster with basic information and some graphics
  - Interested in conference form (includes name and email address)
  - Advertisement/text of conference, to get the word out through different channels (e.g., social media, emails, websites, posters)
- Acceptance emails

- - Different versions depending on each participant e.g., career stage, presentation (talks, poster, attendance), travel support, child care support
  - System for organizing info (names, email addresses, etc.) to create a subset of participants with different requirements/categories
- Pre-/Registration form, which could include
  - Name, location, university
  - Poster or talk title and abstract
  - Career stage
  - Demographics
  - Travel/Dependent Care support
  - Accessibility information
  - Remote participation option (if possible)
- Code of Conduct (examples from IA2 here)
  - For in-person interaction, meeting rooms, and online/Slack chat.
  - Policies for sharing conference content (pictures of presentation slides, pictures of participants, tweeting about discussions, etc.).
- Remote participation agreement or guidelines
- Communication with presenters
  - Information on how to make their talks/posters accessible.
  - Ask permission to share slides, posters, and/or webcast, both internally to attendees and publicly, to the science community.
  - Remote participation (either presenting remotely or permission to share talks with remote participants)

## During the Conference

- Consider different color badges for OC members (so participants can identify them)
- Ensure there is a way for participants to indicate whether or not they wish to be photographed (e.g. with different colored lanyards/visible stickers).
- Welcome packet
  - Documents could be emailed few days earlier and a few printed copies should be available during the conference
  - Packet could include:
    - Badges with pronouns, or pronouns available as stickers
    - Program of the conference
    - Accessibility information
    - Accessible map of the venue identifying bathrooms, exits (both wheelchair accessible and not), rooms, cafeteria etc
    - Map of the local area including restaurants, bars, places to visit
    - Code of Conduct (examples from IA2 here)
    - Conference dinner location, information and directions
    - Transportation information (airport, hotel, venue)
    - Any goodies...

## After the Conference

- Exit survey — emailed to all participants on the last day of the conference OR soon after (example of questions from IA2 [here](here))
    - Questions could be about:
        - [Program structure](Program structure) (Talks, Posters, Discussions, Breaks)
        - Venue facilities (including [Accessibility](Accessibility))
        - [Catering, accommodation, transportation](Catering, accommodation, transportation)
        - [Remote participation](Remote participation) (if used)
        - Conference dinner
    - All forms, questionnaires, etc. should to be tested (by OCs) before they are sent out

# Registration

Our biggest recommendation for this section is transparency; it needs to be clear what decisions you are making regarding registration and why, and these need to be available in a timely manner. There are several goals that you can announce on your website leading up to the conference, including but not limited to:
- What kind of support you can provide to participants
- How many travel/other funding awards you hope to provide
- If using criteria for participation, announce these criteria/goals
- What kind of limitations, if any, on accessibility you have

## Participant Selection

Participant selection may need to happen for a variety of reasons:
- The size of your venue (see [Determining the Spacetime Coordinates](#))
- The focus of your conference (see [Programming](#))
- Making sure the conversation is not dominated by privileged voices

Additionally, you might want to account for known persons to the community that could cause a disruption during, or detract from the primary goals of, the conference. Discussions have been had at previous IA meetings and in the Astro2020 Decadal survey white papers on the status of the workforce about holding the community accountable for "bad behavior" that is, at the least, inconsistent with DEI goals. Finding ways to make sure this is also done for the IA meetings themselves is clearly important.

If you decide to select participants, we recommend making the criteria or goals for participation public before opening registration and allow sufficient time for feedback. One way to do this may be demographics; see [Demographic Information Collection](#). Additionally, we recommend that you release a document outlining your selection method before or at the same time that notifications are sent to applicants so that all interested individuals can understand how the selection was made.

Below is a non-exhaustive list of ways that participants could be selected:
- First-come-first-served cutoff (though could be biased towards time zones, access to computers, etc.)
- Purely randomized selection from a pre-registration pool (this method may result in the exclusion of certain underrepresented groups and is not advised)
- Randomized selection from the pre-registration pool based on participant attributes, demographic or otherwise
- Formal application process in which all applications are reviewed by the OC

Additionally, if selecting participants, whichever method is used for selection, we highly recommend making sure that community leaders and leading thinkers that applied, are invited (this will likely require reaching out to community members to create this list). Just as invited speakers will be separate from this selection, so should active and key leaders in the community.

## Information to be Collected from Participants

The following are some suggested questions you may want to include in the pre-registration or registration. This list is not exhaustive:
- Name
- Email address
- Institute
- Country (and state if US)
- Pronouns (if needed, consider omitting and add options for people to select their pronouns at the conference, e.g. stickers)
- If they would like to present something at the conference
    - Space for title
    - Space for an abstract
    - Type of presentation (talk, poster, panel, or all)
    - If they are not selected to give a talk, do they plan to bring a poster? (if it can be accommodated)
    - Will they be giving a poster/talk with someone else?
    - Optional: justification of talk (see [Programming](#))
    - Electronic vs physical poster
    - Do they give permission for content to be shared? (allow for this to be changed)
- Dependent care needs
- Financial assistance needs
- Accessibility needs/requirements (Include a link to accessibility information in case their needs are already being met. Also include information about who to contact if they wish to confidentially/anonymously communicate this information.)
- Additional Accommodations
- Dietary restrictions/preferences
- Demographic data (if collected; see [Demographic Information Collection](#))
- Career stage
- Career path (industry/academic)
- Allow pictures to be taken during the conference?
- Add space for any additional information they would like to share
- If applicable, add a tick box of acknowledgement that the provided information is for pre-registration only and does not guarantee attendance

Allow for people to update their information before the conference. Depending on how much time is needed to allow for printing of badges and other information, ask registrants to review and update their information (e.g., name, pronouns, dietary restrictions, anything that might have changed), before badges are printed and arrangements finalized.

Badges are important: ensure all information that should be on there is clarified with the printers ahead of the printing deadline. If possible, allow for changes to be made to the badges at the conference (e.g., adding stickers to provide necessary information or reprinting the badge).

# Demographic Information Collection

An important question to ask yourselves when planning this conference, is whether or not to collect demographic information.

Pros:
- You know the demographics of the people who are coming to your conference.
- You have the option of making sure that the accepted attendees represent balanced demographics or focus on selected demographic groups.

Cons:
- Asking for demographic information well can be challenging.
- Some people will not feel comfortable answering demographic questions which can adversely affect selection, if any is made, or statistics.
- There will likely be no agreement from potential participants on whether you asked the right or necessary questions.

To the last point, from our exit survey alone, we had participants in IA2 feel that: our demographics collection was good, that it wasn't specific enough, and that we shouldn't be asking for demographics at all. It likely won't be possible to please all of these viewpoints at once.

## Collecting demographic data

At the most basic level, if collecting demographic information, we recommend making all questions optional. It is important to take into consideration the following:
- This person might be the only one that identifies the way they do where they work and therefore is easily identifiable from the information provided and may not be comfortable answering the question.
- How broad or narrow do you want/need the questions to be?
  - For example, asking if someone identifies as a marginalized race (i.e. not white) could already be identifying enough at a specific institution, but in order to make sure you have representation from people of multiple races/ethnicities you might want to consider asking how they identify. However, this question can also be triggering for some and cause them to not answer at all.

We have found that this can be a very divisive subject where some people can become frustrated if they feel that not enough is asked or, in other cases, that too much is asked. Therefore, in addition to making all of these questions optional, we also encourage you to always have a "prefer not to answer" and write-in sections. Note that allowing freeform answers, while allowing for people to address questions not asked or options not given, can be difficult to process in the case the specific information will not necessarily be an option given to all participants.

If this information is collected, we recommend keeping it separate from any other registration data and only allow access to a minimal number of [OC members](). One way this can be done is by splitting demographic information from the other pre-registration responses and link back to each participant with a unique identifier that is not their name. Communicating how this data is handled is paramount to gaining trust with the registrants and ensuring that their sensitive information will not be compromised.

Consider the following:
- What platform will you use to collect the information? Google may not be the most secure choice, e.g., it can record a person's email address if they are signed into the browser when they take the survey
- Who will have access to the information once it's collected?
- How will you avoid accidentally making the information visible to others?
- What (if any) aggregate data will you release?
- Will you send an email confirmation that you received their responses? Which (if any) of their responses will be copied into that confirmation? This can pose security issues or identity issues if someone registers with a non-private institutional account.
- Share your answers to questions like these in a clear place on relevant form(s).

Asking broad vs specific questions can be one of the most challenging parts. We have given some resources below (which are by no means exhaustive) and recommend future organizers to spend time researching this area and to always pass your questions by *multiple* people from the community that the questions regard. Involving a social scientist or person trained in gathering such information is strongly recommended.

Limited list of recommended resources:
- [Open Demographics]()
- [The Nonbinary Fraction: Looking Towards the Future of Gender Equity in Astronomy]()

# Additional Resources

- The [Committee On INclusion in SDSS (COINS)](#) has several pages dedicated to conference inclusivity:
    - [Meeting Accessibility](#)
    - [Telecon Recommendations](#)
    - [Chairing Guidelines](#)
- [Inclusive Astronomy 2015 Materials](#)
- [Guide to American Astronomical Society (AAS) Meeting Etiquette](#)
- [AAS Working Group on Accessibility and Disability (WGAD) resources and contact information](#)
- [Web Accessibility Initiative guide for accessible presentations](#)
- ["Enhancing Conference Participation to Bridge the Diversity Gap" white paper](#)

# Acknowledgements

We thank everyone that helped us build this document, including the [consulting members](#) of the LOC and the SOC. Thank you to all participants and observers who provided useful feedback before, during, and after the conference. We would also like to thank everyone that chose not to attend on account of the missteps we made along the way, your actions and words have helped us identify issues and learn from them. We hope that by documenting these missteps, future organizers can avoid these same issues to create more inclusive and productive IA meetings. Our experience has shown us that the path to making astronomy truly more diverse, equitable, and inclusive is a challenging but vital one that we must all work on as a community.

# Contact & Referencing this Document

To read the LOC's letter to the community, access an editable version of this document, get information on referencing these recommendations, and find contact information, please visit the [IA2 LOC Recommendations webpage](#).